\documentstyle[amsfonts,amsbsy,12pt]{article}
\def\fmref#1{(\ref{#1})}

\def\vec#1{{\boldsymbol #1}}

\def\i{{\mathrm i}}
\def\lds{|\mkern-2.5mu|}
\def\rds{|\mkern-2.5mu|}

\def\ldc{(\mkern-4mu(}
\def\rdc{)\mkern-4mu)}
\def\dagg{{\setlength{\unitlength}{.09mm}
              \picture(18,30)
              \put(2,20){\line(1,0){16}}
              \put(10,0){\line(0,1){30}}
              \endpicture}}          
\def\ankh{{\setlength{\unitlength}{.09mm}
              \picture(18,30)
              \put(2,23){\line(1,0){16}}
              \put(2,17){\line(1,0){16}}
              \put(7,0){\line(0,1){30}}
              \put(13,0){\line(0,1){30}}
              \endpicture}} 
\def\e{\mbox{e}}
\def\Av#1{\left\langle #1\vphantom{\int}\right\rangle}
\begin{document}
\makeatletter
%%%%%%%%%%%%%%%%%%%%%%%%%%%%%%%%%%%%%%%%%%%%%%%%%%%%%%%%%%%%%%%%%%%%%%
\title{Bose-Einstein correlations of unstable particles
\thanks{Work supported by GSI, 
hypertext link http://www.gsi.de/gsi.html}}
\author{P.A.Henning${}^{a,b,}$
\thanks{Electronic mail: P.Henning@gsi.de, paper mail: c/o 
              Theory Department, GSI}
and Ch.H\"olbling${}^{a}$\\[3mm]
        ${}^{a}$ Theory Department,
         Gesellschaft f\"ur Schwerionenforschung GSI\\
         P.O.Box 110552, D-64220 Darmstadt, Germany\\
         ${}^{b}$Institut f\"ur Kernphysik der TH Darmstadt,\\
         Schlo\ss gartenstra\ss e 9, D-64289 Darmstadt}
\date{\today}
\maketitle
\begin{abstract}
Within a field theoretical formalism suited to treat inhomogeneous
hot quantum systems, we derive the two-particle correlation function
for particles having a spectral width $\gamma$ in the region of
their emission. We find, that this correlation function measures
the radius $R_0$ of the thermal source only in case $\gamma R_0 >1$.
\end{abstract}
\clearpage
%%%%%%%%%%%%%%%%%%%%%%%%%%%%%%%%%%%%%%%%%%%%%%%%%%%%%%%%%%%%%%%%%%%%%%
%\section{Introduction}
In the field of relativistic heavy-ion collisions
the analysis of Bose-Einstein correlations has attracted much 
attention \cite{HBT56,GKW79,P84,BDH94}. 
The general hope is to extract
information about the size of a source radiating pions by studying
their two-particle correlations. These correlations are typical
quantum effects, hence quantum field theory is the proper framework 
to address the problem theoretically. Two types of sources are 
physically interesting: {\em Classical\/} currents and {\em thermal\/}
sources -- the physical reality being a mixture of these cases.

The problem of a free quantum field radiating from
a classical current is exactly solvable \cite[pp.438]{IZ80}, also
at finite temperature \cite{h89coh}. Conversely, it is quite
difficult to find a consistent theoretical description for bosons
radiating thermally from a local ''hot spot''. One reason is,
that such a physical situation corresponds to a non-equilibrium state.
Another reason for this difficulty is, that in thermal states a
breakdown of perturbation theory may occur when it is expressed
in terms of stable particles having zero width \cite{L88}.

With the present paper we attempt to study these two aspects, i.e.,
we address the questions: 1. How are Bose-Einstein correlations affected
by gradients in the temperature distribution, and 2. How are
they affected by a finite lifetime of the bosons. 

In our model, we assume an entirely thermal radiation of pions
from a ``hot spot'', which acquire a nonzero spectral width in 
the region of their generation, 
e.g. by coupling to $\Delta_{33}$-resonances. All thermally
generated pions are propagating out of the hot spot.

This picture is grossly simplified as compared to the physical
reality: At least a substantial fraction of pions generated in
relativistic heavy-ion collisions is not produced thermally,
but arises from the decay of the $\Delta_{33}$-resonance outside
the ``hot spot''. However, on one hand it is legitimate to study 
only a partial aspect of the full problem.
On the other hand, the formalism introduced here is general enough to
be extended to the resonances as well.   

Quantum field theory for non-equilibrium states comes in two flavors:
The Schwinger-Keldysh method \cite{SKF} and thermo field dynamics (TFD) 
\cite{Ubook}. For the purpose of the present paper, we prefer the 
latter method: Apart from its technical elegance, the problem 
of an inhomogeneous temperature distribution has been solved 
explicitly in TFD, up to first order in the temperature gradients 
\cite{h93trans,h94rep}. This solution includes a nontrivial spectral
function of the quantum field under consideration: It employs 
a perturbative expansion in terms of {\em generalized free fields\/}
with continuous mass spectrum \cite{L88}.
%%%%%%%%%%%%%%%%%%%%%%%%%%%%%%%%%%%%%%%%%%%%%%%%%%%%%%%%%%%%%%%%%%%%%%%%
%\section{Outline of the method} 

In ''ordinary'' quantum mechanics,
a statistical state of a quantum system is described by a statistical
operator (or density matrix) $W$, and the measurement of
an observable $\cal{E}$ will yield an average that is calculated as the 
trace of $\cal{E} W$ over the Hilbert space of the system. 
In thermo field dynamics (TFD), this is simplified to the 
calculation of a matrix element
\begin{equation}\label{av3}
\Av{ {\cal E}(t,\vec{x}) } =
\frac{ \ldc 1 \rds \;{\cal E}(t,\vec{x}) \;\lds W\rdc}{
       \ldc 1 \rds  W\rdc}
\;\end{equation}
and the Hilbert space is doubled \cite{Ubook}.
The thermal pion field is described by two scalar field
operators $\phi_x$, $\widetilde{\phi}_x$  and their adjoints
$\phi_x^\star$, $\widetilde{\phi}_x^\star$, with canonical
commutation relations. The field $\phi_x$ is expanded into momentum
eigenmodes: $a_{kl}^\dagg(t)$ creates a pion with
momentum $\vec{k}$ and charge $l=\pm1$, a second set of operators
(commuting with $a^\dagg$, $a$) 
exists for the tildean field $\widetilde{\phi}_x$ \cite{Ubook,h94rep}.

These operators do not excite {\em stable\/} on-shell pions. Rather, they are
obtained as an integral over more general operators
$\xi$, $\widetilde{\xi}$ with a continuous energy parameter $E$: 
\begin{eqnarray}\label{bbg4}\nonumber
\left({\array{r} a_{kl}(t)\\
          \widetilde{a}^\dagg_{kl}(t)\endarray}\right)\;=
& \int\limits_0^\infty\!\!dE\,\int\!\!d^3\vec{q}
  \;\cal{A}^{1/2}_l(E,(\vec{q}+\vec{k})/2)\,
  \left(\widetilde{\cal B}^{-1}_l(E,\vec{q},\vec{k})\right)^\star
  \,\left({\array{r}\xi_{Eql}\\
              \widetilde{\xi}^\dagg_{Eql}\endarray}\right)
  \,\e^{-\mathrm{i} Et} \nonumber \\
\left({\array{r}a^\dagg_{kl}(t)\\
         -\widetilde{a}_{kl}(t)\endarray}\right)^T\;=
& \int\limits_0^\infty\!\!dE\,\int\!\!d^3\vec{q}
  \;\cal{A}^{1/2}_l(E,(\vec{q}+\vec{k})/2)\,
  \left({\array{r}\xi^\ankh_{Eql}\\
           -\widetilde{\xi}_{Eql}\endarray}\right)^T\,
  \widetilde{\cal B}_l(E,\vec{q},\vec{k})
  \,\e^{\mathrm{i} E t}
\;.\end{eqnarray}
The principles of this expansion have been derived in ref. \cite{L88},
$\cal{A}(E,\vec{k})$ is a positive weight function.
For equilibrium states, this function is the spectral function of the
field $\phi_x$. For non-equilibrium systems, the existence
of a spectral decomposition cannot be guaranteed \cite{Ubook}.
We may expect however,
that close to equilibrium the field properties do not change very
much. Thus, with this formalism we study a
quantum system under the influence of small gradients in the
temperature, with {\em local\/} 
spectral function $\cal{A}(E,\vec{k})$. Corrections to such
a picture only occur in second order of temperature gradients
\cite{h93trans,h94rep}. 

A thorough discussion of the $2\times 2$ Bogoliubov matrices was
carried out in ref. \cite{hu92}. For the purpose of the present paper,
we simply state their explicit form as
\begin{equation}\label{gb}
\widetilde{\cal B}_l(E,\vec{q},\vec{k}) = \left( { \array{lr}
   \left(\delta^3(\vec{q}-\vec{k}) + N_l(E,\vec{q},\vec{k})\right)
            \;\;\;& -N_l(E,\vec{q},\vec{k}) \\
   -\delta^3(\vec{q}-\vec{k})     & \delta^3(\vec{q}-\vec{k})
   \endarray} \right)
\;,\end{equation}
where $N(E,\vec{q},\vec{k})$ is the Fourier transform of a space-local
Bose-Einstein distribution function
\begin{eqnarray}\nonumber \label{nloc}
N_l(E,\vec{q},\vec{k})& = & \frac{1}{(2\pi)^3}\;
        \int\!\!d^3\vec{z}\,\e^{-\mathrm{i}
  (\vec{q}-\vec{k})\vec{z} }\,n_l(E,(\vec{q}+\vec{k})/2,\vec{z})\\
n_l(E,(\vec{q}+\vec{k})/2,\vec{z}) & = &  
  \frac{1}{\e^{\beta(\vec{z}) (E-\mu_l(\vec{z}))}-1}
\;.\end{eqnarray}
The $\xi$-operators have commutation relations
\begin{equation}\label{difc}
\left[\xi_{Ekl},\xi^\ankh_{E^\prime k^\prime l^\prime}\right]=
  \delta_{ll^\prime}\,
  \delta(E-E^\prime)\,
\delta^3(\vec{k}-\vec{k}^\prime)
\;.\end{equation}
Similar relations hold for the $\widetilde{\xi}$ operators, 
all other commutators vanish, see \cite{L88}. They act on the
''left'' and ''right'' statistical state according to
\begin{equation}\label{tscc}
\xi_{Ekl}\lds W \rdc = 0, \;\;
\widetilde{\xi}_{Ekl}\lds W \rdc =0,\;\;
\ldc 1\rds \xi^\ankh_{Ekl} = 0 ,\;\;
\ldc 1\rds \widetilde{\xi}^\ankh_{Ekl} = 0\;\;\;\forall\,E,\vec{k},l=\pm1
\;. \end{equation}
With these rules, all bilinear expectation values can be
calculated exactly. Higher correlation functions have
a perturbative expansion in the spectral function.
%%%%%%%%%%%%%%%%%%%%%%%%%%%%%%%%%%%%%%%%%%%%%%%%%%%%%%%%%%%%%%%%%%%%%
%\section{Two-particle correlation function}

Of these, we are interested in the two-particle correlation function,
which is the probability to find 
in the system a pair of pions with momenta $p$ and $q$: 
\begin{equation} \label{cfc}
c_{ll^\prime}(\vec{p},\vec{q})=\frac{\Av {a^\dagg_{pl}(t) 
 a^\dagg_{ql^\prime}(t)
            a_{ql^\prime}(t) a_{pl}(t)}}{
   \Av{a^\dagg_{pl}(t) a_{pl}(t)}\;
   \Av{a^\dagg_{ql^\prime}(t) a_{ql^\prime}(t)}}
=1+\delta_{ll^\prime}
\frac{\cal{F}_2(\vec{p},\vec{q})\cal{F}_2(\vec{q},\vec{p})
      }{\cal{F}_1(\vec{p}) \cal{F}_1(\vec{q})}
\;,\end{equation}
For simplicity, we abbreviate the mean momentum of this pair by
$\vec{Q}= (\vec{q}+\vec{p})/2$. Using the above rules of 
thermo field dynamics, the functions $\cal{F}_1$ and $\cal{F}_2$ 
are calculated as
\begin{eqnarray} \label{cfcf}
\cal{F}_1(\vec{p}) & = &
      \int\limits_{0}^{\infty}\!dE\int\!d^3\vec{z}\,\cal{A}_l(E,\vec{p})
      n_{l}(E,\vec{p},\vec{z})\nonumber\\
\cal{F}_2(\vec{p},\vec{q}) & = &
      \int\limits_{0}^{\infty}\!dE\int\!d^3\vec{z}\,
      \left(\cal{A}_l(E,\vec{p})
            \cal{A}_l(E,\vec{Q})\right)^{\frac{1}{2}}\,
      \mbox{e}^{\i(\vec{p}-\vec{q})\vec{z}}\,
      n_{l}(E,\vec{Q},\vec{z})
\;. \end{eqnarray}
Before we use the above expression to obtain numerical results,
we perform an expansion of $\cal{F}_2$ around the mean momentum $\vec{Q}$.
This is consistent with the restriction, that the spectral function
acquires corrections in second order of the gradients:
\begin{eqnarray} \label{cfc2} \nonumber
\cal{F}_2(\vec{p},\vec{q}) & = &  \cal{F}^0_2(\vec{p},\vec{q}) \\  
\nonumber 
&+&
   \int\limits_{0}^{\infty}\!dE\int\!d^3\vec{z}\,
   \mbox{e}^{\i(\vec{p}-\vec{q})\vec{z}}\,\left(
   \frac{\i}{2} \nabla_{\vec{Q}}\cal{A}_l(E,\vec{Q})\,
   \nabla_{\vec{z}} n_{l}(E,\vec{Q},\vec{z})\right) + \cal{O}(\nabla^2_z n)\\
\cal{F}^0_2(\vec{p},\vec{q}) &=&  
\int\limits_{0}^{\infty}\!dE\int\!d^3\vec{z}\,
   \mbox{e}^{\i(\vec{q}-\vec{p})\vec{z}}\,
   \cal{A}_l(E,\vec{Q})\,
   n_{l}(E,\vec{Q},\vec{z})
\;.\end{eqnarray} 
The first part of this expansion is, 
apart from the folding with the spectral function, 
also obtained in other calculations of the correlator \cite{GKW79,P84,BDH94}.
This {\em standard\/} expression for the correlation function therefore is
\begin{equation}\label{cfc3}
\overline{c}_{ll^\prime}(\vec{p},\vec{q})
  =1+\delta_{ll^\prime}
\frac{\cal{F}^0_2(\vec{p},\vec{q})\cal{F}^0_2(\vec{q},\vec{p})
      }{\cal{F}_1(\vec{p}) \cal{F}_1(\vec{q})}
\;,\end{equation}
and our result for $c_{ll^\prime}(\vec{p},\vec{q})$ 
differs from $\overline{c}_{ll^\prime}(\vec{p},\vec{q})$
in first order of gradients in the distribution function.
For the purpose of generalizing our result it is worthwhile to note
that the gradient term in \fmref{cfc2} is just one half of the 
Poisson bracket of $\cal{A}$ and $n$
\cite{h93trans}.
%%%%%%%%%%%%%%%%%%%%%%%%%%%%%%%%%%%%%%%%%%%%%%%%%%%%%%%%%%%%%%%%%%%%%
%\section{Results and discussion}

We have calculated these
correlation functions with a simple parameterization
of the pion spectral function, 
\begin{equation}
  \cal{A}_l(E,\vec{p})=\frac{2E\gamma}{\pi}
  \frac{1}{(E^{2}-\Omega_{p}^{2})^{2}+4E^{2}\gamma^{2}}
\end{equation}
where $\Omega_{p}=\sqrt{m_{\pi}^{2}+\vec{p}^{2}+\gamma^{2}}$ and
$m_\pi$ = 140 MeV. This 
parameterization has been motivated and related to a more serious
field theoretical approach in ref. \cite{h93pion}. To 
gain information about the {\em maximal\/} influence exerted
by the occurrence of a nonzero spectral width, we studied
only the case of an energy and momentum independent $\gamma$
equal for both charges. The temperature distribution was taken as
radially symmetric gaussian
\begin{equation} \label{tempd}
 T(\vec{z})=T(r) = T_0 \exp\left(-\frac{r^{2}}{2R_0^{2}}\right)
\;,\end{equation}
with chemical potential $\mu=0$ and $R_0$ = 5 fm.
The local equilibrium pion distribution
for a given momentum $\vec{k}$ is obtained by folding $n$ with the
spectral function. Hence, the mean radius of this particle
distribution function acquires a $\gamma$-dependence. We define
the rms radius {\em orthogonal\/} to the direction of $\vec{Q}$ as
\begin{equation}\label{rav}\nonumber
R_n = \sqrt{\frac{I_2}{I_0}} \;\;\;\;\;\;\;
I_j = \int\limits_0^\infty \!\!dr\,r^j\,\int\limits_0^\infty
  \!\!dE\,\cal{A}(E,\vec{k})\,\left(\mbox{e}^{E/T(r)}-1\right)^{-1}
\;.\end{equation}
Note, that $R_n$ is {\em not\/} the 3-dimensional
rms radius of the distribution function (which would be $I_4/I_2$).
Rather, $R_n$ is half the product of angular
diameter and distance between detector and source.
A constant temperature over a sphere of radius $R_0$
would yield an $R_n = R_0/\sqrt{3}$, while
its 3-D rms radius is $R_0 \sqrt{3/5}$.

The correlation functions, $c_{ll^\prime}(\vec{p},\vec{q})$ 
calculated according to \fmref{cfc} and 
$\overline{c}_{ll^\prime}(\vec{p},\vec{q})$ 
as defined in eqn. \fmref{cfc3} then may be fitted by a gaussian form, i.e.,
\begin{equation}\label{gau}
c_{ll^\prime}(\vec{p},\vec{q}) \approx 
1 + \exp\left( -R^2 (\vec{p}-\vec{q})^2 \right)
\;\end{equation}
and similarly for $\overline{c}_{ll^\prime}(\vec{p},\vec{q})$ 
with parameter $\overline{R}$.
We assume for our conclusion, that $c_{ll^\prime}(\vec{p},\vec{q})$ is the
correlation function measured experimentally.
 
In the figure, we have plotted the two fit parameters $R$, $\overline{R}$ 
and $R_n$ as function of $\gamma$. The principal
result of the calculation is, that for {\em small\/} values
of $\gamma$ the correlation function $c_{ll^\prime}(\vec{p},\vec{q})$ becomes
{\em narrower\/} in momentum space than 
$\overline{c}_{ll^\prime}(\vec{p},\vec{q})$.
Consequently, the {\em measured\/} correlation radius $R$
is always larger than $\overline{R}$ as expected from the function
$\overline{c}_{ll^\prime}(\vec{p},\vec{q})$. 
The deviation is such that
for small enough $\gamma$, $R \approx \overline{R} + 1/\gamma$.
For larger $\gamma$, the small differences between $R$, $\overline{R}$
and $R_n$ may be attributed to our use of a gaussian temperature distribution:
$n(T(r))$ is not stricly gaussian,  only in the (unphysical) limit
$\gamma\rightarrow\infty$ one reaches $R=\overline{R}=R_n=R_0/\sqrt{2}$.
%%%%%%%%%%%%%%%%%%%%%%%%%%%%%%%%%%%%%%%%%%%%%%%%%%%%%%%%%%%%%%
\begin{figure}[t]
\setlength{\unitlength}{1mm}
\begin{picture}(150,90)
\put(-3,73){$R$ [fm]}
\put(90,43){$R_0/\sqrt{2}$ }
\put(60,13){$R_0/\sqrt{3}$ }
\put(120,7){$\gamma$ [MeV]}
\end{picture}
%%
%% dvips
\includegraphics{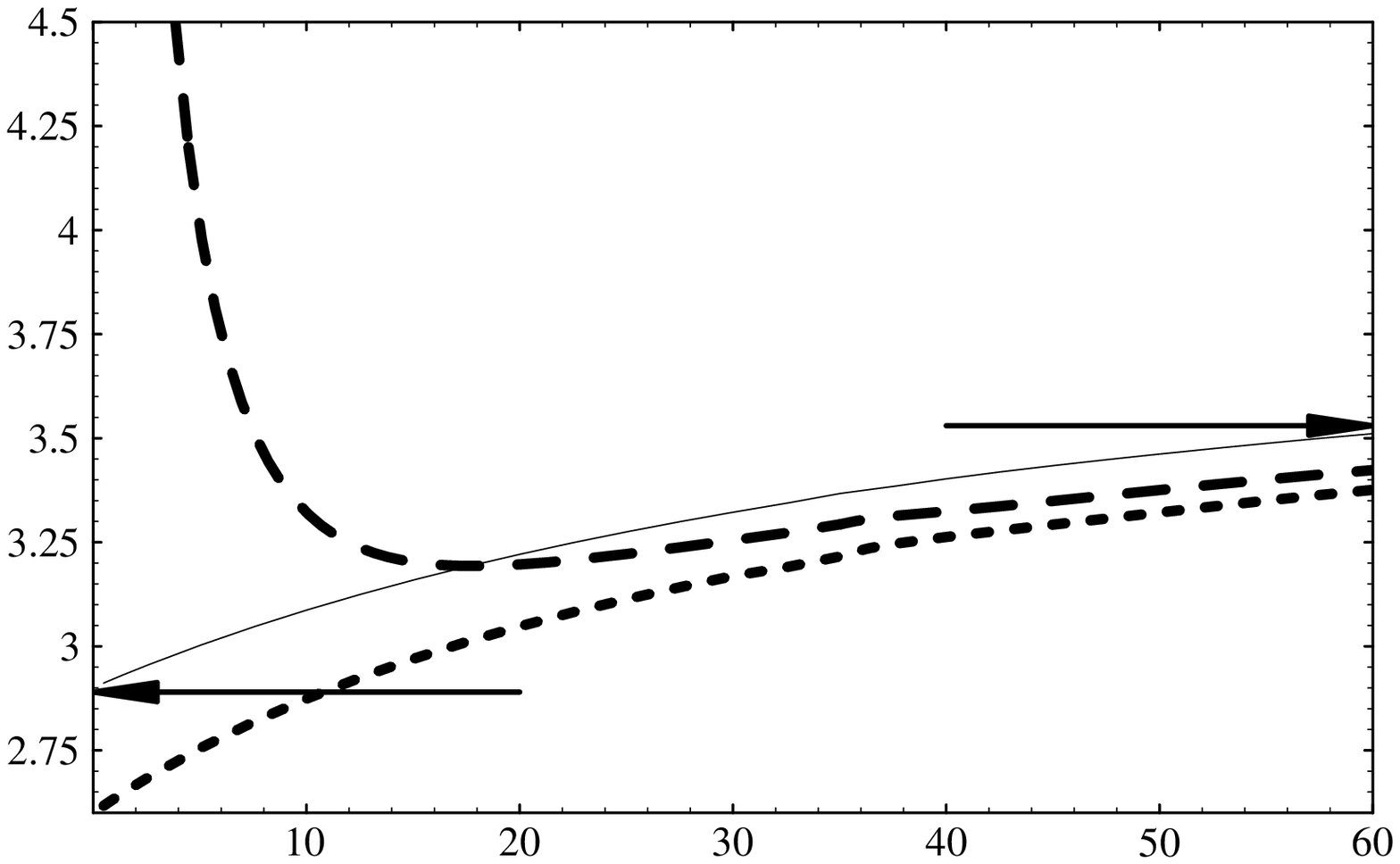}
\caption{Correlation radius of a pion ``hot spot'' with 
temperature $T_0$=100 MeV}
Long dashed line: Gaussian fit to 
$c_{ll^\prime}(\vec{p},\vec{q})$ according to \fmref{cfc} \\
Short dashed line: Gaussian fit to 
  $\overline{c}_{ll^\prime}(\vec{p},\vec{q})$ according to \fmref{cfc3}. \\
Continuous line: mean radius of the thermal distribution function,
eqn. \fmref{rav}.\\
\hrule
\label{fig1}
\end{figure} 
%%%%%%%%%%%%%%%%%%%%%%%%%%%%%%%%%%%%%%%%%%%%%%%%%%%%%%%%%%%%%%%

Before we interpret this result, we have to admit that
our calculation is very crude: Neglecting
energy and momentum dependence of the $\gamma$ in the
spectral function can be a first step only. Also, in a more realistic
calculation the partially {\em coherent\/} production of pions
would have to be taken into account, forcing 
$c_{ll^\prime}(\vec{p},\vec{p})<2$.
 
Within this limitation however we feel safe when stating the following 
answer to the questions raised in the introduction: A finite lifetime 
or nonzero spectral width $\gamma$ of the bosons is essential,
if one wants to infer the {\em thermal source radius\/} $R_0$ 
from correlation measurements. To be more precise, 
only for $\gamma R_0 \ge 1$ the correlation function measures 
the mean radius of the particle distribution function. 

This result is also in agreement with our view of the {\em equilibration
process\/}: The equilibration rate of a distribution function is, to
lowest order, given by the spectral width of the particle \cite{hu92}.
Consequently, a very small $\gamma$ corresponds to a system that does
not equilibrate -- hence the correlation function approaches
the ``quantum limit'' 
$c_{ll^\prime}(\vec{p},\vec{q})\rightarrow 1 
+ \delta_{ll^\prime}\delta_{pq}$,
and the correlation radius obtained by a gaussian fit becomes infinite. 

Also, for a given energy, $1/\gamma$ is a 
measure for the spatial size of the pion ``wave packet'', which must
be smaller than the object to be resolved. 
In {\em stars\/} emitting photons
having only their thermal width $\gamma\approx \alpha T\approx 0.5/137$ eV, 
the condition $\gamma R_0 \gg 1$ is always satisfied. 
For the typical thermal pion sources
occurring in relativistic heavy-ion collisions however,
this criterion can be {\em violated\/}: Pions of 100 MeV momentum
in nuclear matter have an effective $\gamma$ of only a few MeV 
\cite{h93pion}.

We conclude, that in the Hanbury Brown-Twiss analysis of relativistic 
heavy-ion collisions one may measure {\em correlation radii\/},
i.e., by gaussian fits to the correlation function, which are
much larger than the actual radius of the thermal source.
Most certainly it is {\em not\/} possible to infer the size
of a thermal source by correlation measurements, when the
spectral function of the measured bosons in the region of
their generation is unknown. As a rule of thumb we suggest, that
pion correlation functions should be measured for momenta of
$|\vec{Q}| \approx$ 350 MeV, since then the mixing of pions with 
$\Delta_{33}$-resonance/nucleon-hole excitations is largest.

As a final note we emphasize again, that in common calculations 
of the correlation function one has to introduce ad-hoc
random phases between several classical sources and then obtains
the ``standard'' correlator 
$\overline{c}_{ll^\prime}(\vec{p},\vec{q})$. Instead, we
relied on a proper field theoretical treatment correct up to first order in
gradients of the temperature. We found, that the non-equilibrium character
of the system must be taken serious when calculating the correlation 
function.
%%%%%%%%%%%%%%%%%%%%%%%%%%%%%%%%%%%%%%%%%%%%%%%%%%%%%%%%%%%%%%%%%%%%%

\end{document}